\documentclass[prl,twocolumn,graphicx,amssymb,showpacs]{revtex4}
\usepackage{graphicx}
\begin{document}

\title{On the Role of Potentials in the Aharonov-Bohm Effect  }

\author{Lev Vaidman}

\affiliation{ Raymond and Beverly Sackler School of Physics and Astronomy\\
 Tel-Aviv University, Tel-Aviv 69978, Israel}

\begin{abstract}
There is a consensus today that the the main lesson of the Aharonov-Bohm effect is that a picture of electromagnetism based on the
local action of the field strengths is not possible in quantum mechanics.
Contrary to this statement it is argued here that when the source of the electromagnetic potential is treated in the framework of quantum theory, the Aharonov-Bohm effect can be explained without the notion of potentials. It is explained by local action of the field of the electron on the source of the potential. The core of the Aharonov-Bohm effect is the same as  the core of quantum entanglement: the quantum wave function describes all systems together.
  \end{abstract}

\maketitle

Before the  Aharonov-Bohm effect \cite{AB59} (AB) was discovered, the general consensus was that particles can change their motion only due to fields at their locations,  fields which were created by other particles. The main revolutionary aspect of the AB effect was that this is not generally true, and that in certain setups two   particles,  prepared in identical states, move in  the same fields but end up in different final states. In particular, the electromagnetic field can vanish at every place where the electron has been, yet the electron motion is affected by the electromagnetic interaction. The AB effect states that the motion of an electron is completely defined by the potentials in the region of its motion and not just by the fields.
 The potentials depend on the choice of gauge, which cannot affect the motion of  particles, but there are gauge invariant properties of the potentials (apart from the fields) that specify the motion of  particles.
 The validity and the meaning of the AB effect has been extensively discussed \cite{Fu60,Pe61,AB61,Lie65,Boy73,Wu75,Boc78,Roy80,Pes81,Gre81,Ola85,Pes89,AKN09,Wal10}.  I argue that there is an alternative to commonly accepted mechanism which leads to  the effect,  and that we might change our understanding of the nature of physical interactions back to that of the time before the AB effect was discovered. The quantum wave function changes due to local actions of fields.

The discussion will be on the level of  gedanken experiments, without questioning the feasibility of such experiments in today's laboratory.
Consider  a Mach-Zehnder interferometer for electrons tuned in such a way that the electron always ends up in  detector $B$, see Fig.~1. We can change the electric potential in one arm of the interferometer  such that there will be no electromagnetic field at the location of the wave packets of the electron but, nevertheless, the electron will change its behavior and sometimes (or it can be arranged that always) will end up in   detector $A$. This is the electric AB effect. Alternatively, in the magnetic AB effect, the interference picture can be changed due to  a solenoid inside the interferometer which produces no electromagnetic field  at the arms of the interferometer.

\begin{figure}[b]
  \includegraphics[height=5.0cm]{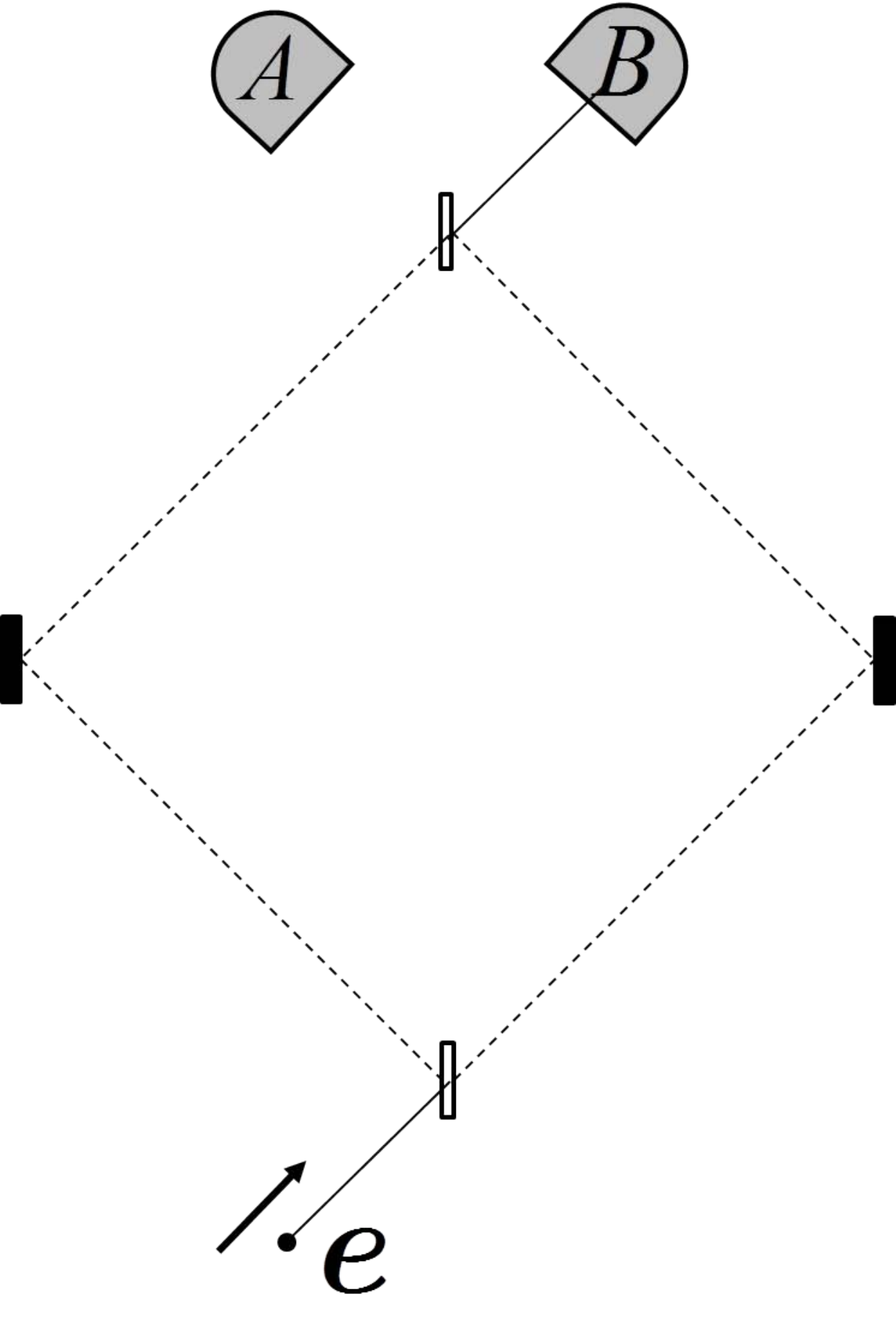}\\
      \caption{{\bf Mach-Zehnder interferometer with electron as a test bed of the AB  effect.} Introducing relative electric potential between the arms of the interferometer or introducing a solenoid inside the interferometer spoils the destructive interference in detector $A$.  } \label{7}
\end{figure}

Let us start our analysis with the electric AB effect.  In the original proposal, the potential was created using conductors, capacitors etc. While those are closer to a practical realization of the experiment, a precise theoretical description of such devices is difficult.  I consider, instead,   two   charged particles,  the fields of which cancel  at the location of the electron.

For simplicity of presentation, instead of the Mach-Zehnder interferometer,  I shall  consider a one dimensional interferometer, see Fig.~2. (In fact, for an observer moving with a constant velocity in a perpendicular direction, this interferometer looks very much like the one described in Fig. 1.) The electron wave packet starts moving to the right toward a  barrier which transmits and reflects equal weight wave packets toward  mirrors $A$ and $B$. After reflection from the mirrors, the wave packets split again on the barrier. The interferometer is tuned in such a way that the there is a complete destructive interference toward mirror $A$,  and the electron reaches   mirror $B$ with certainty.

\begin{figure}[t]
  \includegraphics[height=5.6cm]{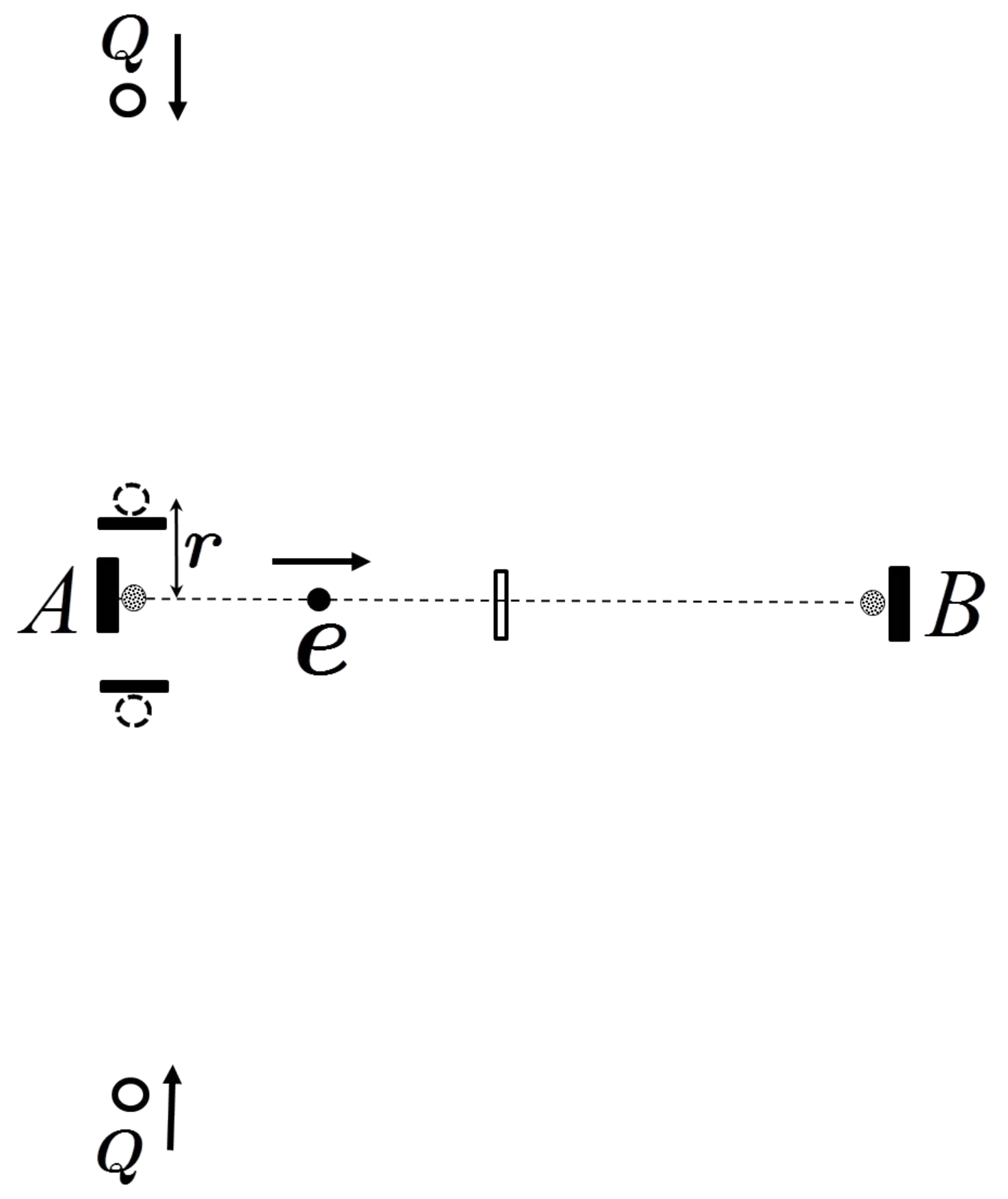}\\
      \caption{{\bf A realization of the electric AB effect.} Identical charges brought symmetrically to the electron wave packet in the left arm of the interferometer create potential for the electron without creating electric field at its location. } \label{2}
\end{figure}

Another modification (the sole purpose of which is simplicity of the quantitative analysis of the experiment)  is designing  a special mirror for the electron which makes it spend  a long  time $\tau$ near it.  For this purpose we introduce an interaction between the electron  and the mirror with  potential energy as a function of the electron distance from the mirror  shown in Fig.~3. It goes to infinity at the surface of the mirror, smoothly becomes a constant value $V$  at $x\in(0,d)$, and smoothly goes to zero for $x>d$. The energy of the electron   is only slightly higher than $V$.  The dimensions of the interferometer are much larger than $d$
 and  we state that the electron is near the mirror when $x\in(0,d)$.

\begin{figure}[b]
  \includegraphics[height=3.5cm]{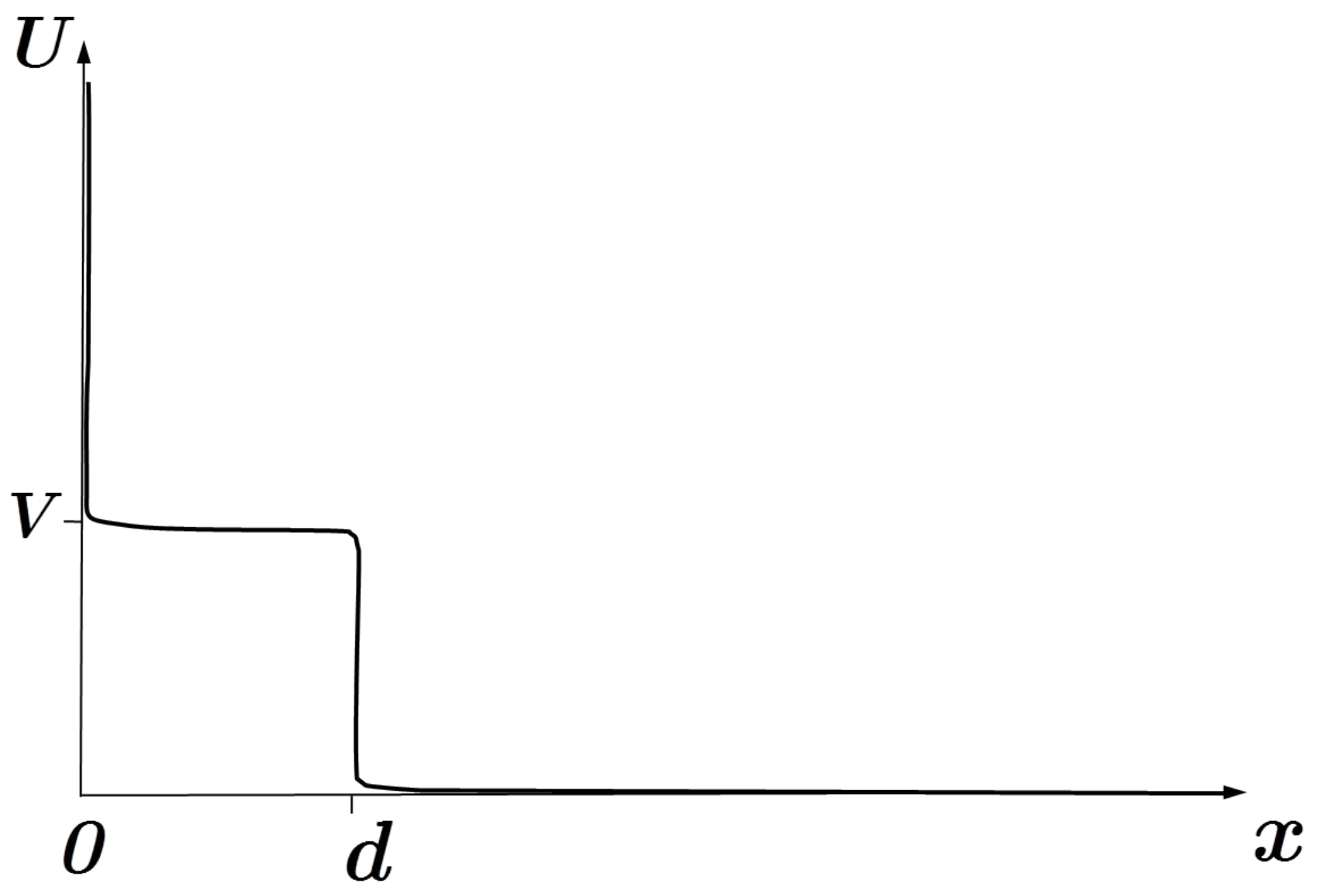}\\
      \caption{{\bf The potential of the mirror forces.} The potential energy of the particle as a function of its distance from the mirror. The particle with an energy slightly higher than $V$ spends long time near the mirror. } \label{3}
\end{figure}

The source of the AB potential will be  two particles of mass $M$ and charge $Q$  placed  symmetrically on the perpendicular axis at equal large distances from mirror $A$. They have equal initial velocities toward the location of mirror $A$. At equal distance $r$ from the mirror, the charged particles bounce back due to other similarly designed mirrors, which make the charges spend a  time $T$ near these mirrors. We choose  $T < \tau $, so that the charges $Q$ are near their respective mirrors during the time the electron's wave packets are near their mirrors.
 We then can approximate the potential that the electron in the left arm feels as $\frac{-2eQ}{r}$ for the  time $T$. Indeed, when the charges are far away, their potential can be neglected, and the time  the charges travel toward and from the mirror is much smaller than $T$. Thus,  the phase difference between the two wave packets of the electron is:
 \begin{equation}\label{ABpha}
 \phi_{AB}=\frac{-2eQT}{r\hbar}.
 \end{equation}
The electron does not feel an electric field at any place where its wave packet passed, but it exhibits an interference pattern which is different from the pattern obtained in such an experiment by a neutral particle.

How can this result be understood if we consider all particles? The quantum state of the composite system is  a superposition of two product states which I name branches. In the first one, the wave packet of  the electron is on the left and in the other, it is on the right. The   energy in the left branch is equal to the energy  in the right branch, so energetic considerations cannot explain the phase difference. The electron does not experience any electric force, so the electron's wave packets are not shifted and thus  cannot provide an explanation of the effect. The charges $Q$, however, do feel different forces in different branches. Thus, their wave packets in the left branch are slightly shifted relative to their wave packets in the right branch.

Let us calculate the shift of position of the wave packet of one of the two $Q$  charges due to its electromagnetic interaction with the electron. The shift is developed during the time $T$ when this charge is near its mirror.  The interaction with the electron leads to a small perturbation in the motion of the charge and, since $d\ll r$,    the velocity of the charge  during this time, $v$, can be considered to be constant.
The change in the kinetic energy of the charge due to its interaction with the electron  allows us to find the change in its velocity and thus the shift $\delta x$ we are looking for:
 \begin{equation}\label{delta}
     \frac{-eQ}{r} =\delta(\frac{Mv^2}{2})\simeq Mv\delta v~~~~~\Rightarrow~~~~\delta x= \frac{-eQT}{Mvr}.
\end{equation}

To observe the interference in the AB experiment, this shift should be  much smaller than the position uncertainty of the charges. The de Broglie wavelength of the charge  $\lambda=\frac{h}{Mv}$. Both charges $Q$ are shifted in the same way creating the AB phase:
$
     2\frac{\delta x}{\lambda}2\pi =\phi_{AB}.
$

The entanglement between the electron and the charges, which could have been created if the uncertainty in the velocity of the charges when they are near their mirrors is smaller than $\delta v$,  disappears when the charges $Q$ travel back. Note however, that if, contrary to our assumption, the position uncertainty of the charges is smaller than $\delta x$, then the entanglement  will remain and will lead to decoherence  washing out the AB effect.

Let us turn now to the magnetic AB effect.  I will show that the AB effect arises from  different  shifts of the wave packets of the source which feels different local electric fields created by the left and the right wave packets of the electron.

Consider the following setup. The solenoid consists of two  cylinders of  radius $r$, mass $M$, large length $L$  and  charges $Q$ and $-Q$ homogenously  spread  on their surfaces. The cylinders   rotate  in  opposite directions with surface velocity $v$. The electron encircles the solenoid with velocity $u$ in superposition of being in  the left and in the right sides of  the circular trajectory  of radius $R$, see Fig.~4.

\begin{figure}[t]
  \includegraphics[height=6.8cm]{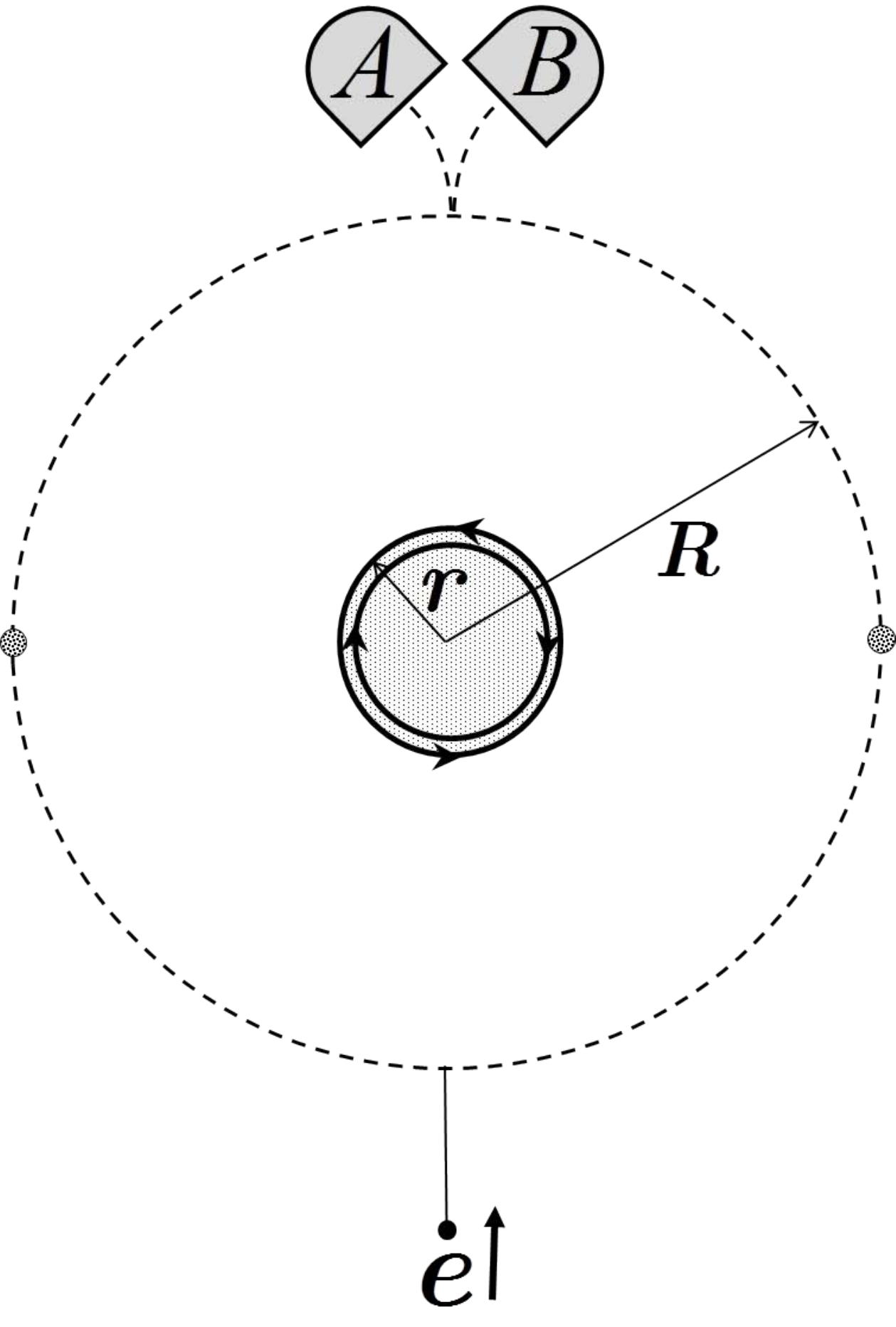}\\
      \caption{{\bf The magnetic AB effect.} The electron wave packet coming directly toward the solenoid splits into a superposition of two wave packets which encircle the solenoid from two sides and come out almost in the same direction, interfering toward detectors $A$ and $B$. } \label{4}
\end{figure}

The flux  in the solenoid due to the two cylinders is:
\begin{equation}\label{flux}
\Phi =2~ \pi r^2 ~ \frac{4\pi}{c} \frac{Qv}{2\pi r L}  =\frac{4\pi Q v r}{cL}.
\end{equation}
Thus, the AB phase, i.e., the change in the relative phase between the left and the right wave packets due to the electromagnetic interaction is:
\begin{equation}\label{ABpha1}
\phi_{AB}=\frac{  e \Phi}{c\hbar}=\frac{4\pi e Q v r}{c^2L\hbar}.
\end{equation}

 To simplify the alternative calculation based on direct action of the electromagnetic field, we assume $r \ll R \ll L$. Before entering the circular trajectory, the electron moves toward the axis of the solenoid and thus it  provides zero total flux through any cross section of the solenoid. During its motion on the circle, the magnetic flux through a cross section of the solenoid at distance $z$  from the perpendicular drawn from the electron is:
\begin{equation}\label{phitheta}
\Phi(z)=\frac{\pi r^2 e u R }{c (R^2+z^2)^{\frac{3}{2}}}.
\end{equation}
 When the electron enters one arm of the circle, it changes the magnetic flux and  causes an electromotive force on the charged solenoids  which changes their angular velocity. In order to calculate the change in the velocity of the surface of the cylinder we have to integrate the impulse exerted on all thin slices of the cylinder. For simplicity, I consider here the surface motion  as a linear motion. The contribution of a slice with an infinitesimal charge $dQ$ to the impulse is  $\frac{\Phi(z) dQ}{c2\pi r}$ and integration over the slices  yields:
\begin{equation}\label{deltav}
\delta v= \frac{1}{M}  \int_{-\frac{L}{2}}^{\frac{L}{2}} \frac{\pi r^2 e uR }{c^2 (R^2+z^2)^{\frac{3}{2}}}\frac{1}{2\pi r}\frac{Q}{ L}  dz\simeq\frac{uQer}{c^2MRL}.
\end{equation}
Then, the shift of the wave packet of a cylinder  during the motion of the electron is:
\begin{equation}\label{deltax}
 \delta x=\delta v \frac{\pi R}{u}=\frac{\pi Qer}{c^2ML}.
\end{equation}
 The relevant wavelength of de Broglie wave of each cylinder is $\lambda= \frac{h}{Mv}$. For calculating the AB phase we should take into account that both cylinders are shifted and that they are shifted (in opposite directions) in both branches. This leads to a factor 4 and provides the correct expression for the AB phase:
$
 4\frac{ \delta x}{\lambda}2\pi=\phi_{AB}.
$

 If the   uncertainty in the velocity of the cylinders  is smaller than  $\delta v$, then, during the electron circular motion, the electron and the cylinders  become entangled. But when the electron leaves the circular trajectory, it exerts an opposite  impulse on the cylinders and this entanglement  disappears.

I have explained both electric and magnetic AB effects through actions of local fields on the quantum wave function. The electron in  states $|L\rangle_e$ and $|R\rangle_e$  causes, via action of its electromagnetic field, different evolutions for the quantum state of the source: $|\Psi_L\rangle_S$ and $|\Psi_R\rangle_S$.
 The total wave function of the electron and the source is
 \begin{equation}\label{psitotal}
 \frac{ 1}{\sqrt 2} \left (|L\rangle_e|\Psi_L\rangle_S+|R\rangle_e|\Psi_R\rangle_S\right ).
\end{equation}
During the evolution, the source states  $|\Psi_L\rangle_S$ and $|\Psi_R\rangle_S$ might become orthogonal, or mostly differ only in their phase, but at the end of the process, the states of the source  are identical except for the AB phase. Thus, the total wave function becomes
 \begin{equation}\label{psitotalF}
  \frac{ 1}{\sqrt 2} |\Psi\rangle_S \left(|L\rangle_e+e^{i\phi_{AB}}|R\rangle_e \right),
\end{equation}
and the AB phase can be observed in  the electron interference experiment.

The celebrated manifestation of a quantum wave function for a combined system is the nonlocal correlations which are generated by entangled states. The AB effect is conceptually different, since it can appear even if in the state (\ref{psitotal}) there is  almost no entanglement at all times.

One might wonder why, instead of performing exact calculations in the framework of quantum mechanics,  I consider  particles and  cylinders pushed by fields in the framework of classical mechanics and then use  the correspondence  principle to calculate the shifts of the quantum wave packets of particles and cylinders. I have to follow this path because  the standard formulation of quantum mechanics and the Schr\"{o}dinger equation in particular, are based on potentials.
I hope that a general formalism of quantum mechanics based on local fields will be developed. It will provide a solution to the problem of motion of a quantum particle in a force field even if there is no potential from which it can be derived. Meanwhile my assertion provides one useful corollary: If the fields vanish at locations of all particles then these fields yield no observable effect.

Let us test this corollary. Consider a modification of the electric AB effect described above in which the charges $Q$ do not automatically perform their motion toward mirror $A$ and back, but only when the electron on the path $A$ triggers this motion, i.e., only in the left branch. I choose a particular value of the charge of  the external particles, $Q=4e$ for which the total electric field at the location of each particle created by other particles is zero. Neither the electron, nor the charges $Q$ feel an electromagnetic  field  in any  of the branches. My assertion is that there will be no AB effect in this setup, in spite of the fact that the electron of  the left branch has an electric potential, while the electron of the right branch  has not. The original treatment of the  AB effect is invalid  since we do not have here a motion of an electron in a classical electromagnetic field, but a treatment of the problem using  ``private potential'' created by induced charges \cite{Kau11}  shows that indeed there is no AB effect in this case.

I believe that we can find an explanation of the kind  presented above for any  model of the AB experiment. However, the pictorial explanation of the creation of relative phase due to spatial shifts of wave packets  disappears when  we go beyond the physics of moving charges. We can replace the charged cylinders by a line of polarized neutrons producing magnetic flux due to quantum spins. In this case there is no spatial shift of wave packets. I am not aware of any pictorial explanation of the change of the phase of the spin state of the neutron, but contrary to the phase of the electron in the standard approach to the AB effect, the phases of neutrons  are  changed locally due to the magnetic field of the electron.  This is also an explanation of the Aharonov-Casher (AC) effect \cite{AC}: the local electric field acting on the moving neutron is responsible for appearance of the AC phase.   Note, however, that it does not lead to a  classical lag of the center of mass of the neutron \cite{ACBo,APV}.

I have not presented  a general proof  that in order to have  an observable effect, the particles must pass through regions of nonzero fields. Rather, what I have shown is that the setups of electric and magnetic AB effects do not contradict this assertion. Note, however, that the last example  in which there is an electric field almost everywhere except the locations of the particles and this field causes no effect, strongly  supports my claim.

Since the electromagnetic potential at any point along the trajectory of the electron can be gauged away, the standard approach to the AB effect leads to a paradoxical, in my view, nonlocal feature of quantum mechanics: the AB phase which has observable manifestation is acquired inside the interferometer in spite of the fact that there is no  particular place or time where this happens. I have shown that this peculiarity disappears when all relevant parts of the system are considered: the phase is gradually acquired by the source of the electromagnetic potential.

 This result does not question the validity of the AB effect and does not diminish the importance of its numerous applications. It removes,  however, conceptual claims associated with the AB effect regarding non-locality and the meaning of potentials.
The AB effect does not prove that the evolution of a composite system of charged particles  cannot be described completely by fields at locations of all particles. The potentials might be just a useful auxiliary mathematical tool after all.

I thank  Noam Erez, Yaron Kedem, Shmuel Nussinov and Philip Pearle for useful discussions. This work has been supported in part by the Binational Science Foundation Grant No. 32/08 and
 the Israel Science Foundation  Grant No. 1125/10,

\end{document}